\documentclass[twocolumn,showpacs,preprintnumbers,amsmath,amssymb,prl,tightenlines,final]{revtex4}
\usepackage{amsfonts,amssymb,amsmath,times}
\usepackage{hyperref,graphics} 

\usepackage{graphicx}
\usepackage{dcolumn}
\usepackage{bm}
\usepackage{epsfig}
\usepackage{longtable}

\input epsf

\begin{document}

\title{Efficient atomic self-interaction correction scheme for non-equilibrium quantum transport} 
\date{\today} 
\author{C. Toher and S. Sanvito}
\email{sanvitos@tcd.ie}
\affiliation{School of Physics, Trinity College, Dublin 2, Ireland}

\begin{abstract}
Density functional theory calculations of electronic transport based on local exchange and
correlation functionals contain self-interaction errors. These originate from the interaction of 
an electron with the potential generated by itself and may be significant in  metal-molecule-metal
junctions due to the localized nature of the molecular orbitals. As a consequence, insulating
molecules in weak contact with metallic electrodes erroneously form highly conducting junctions,
a failure similar to the inability of local functionals of describing Mott-Hubbard insulators. 
Here we present a fully self-consistent and still computationally undemanding self-interaction 
correction scheme that overcomes these limitations. The method is implemented in  
the Green's function non-equilibrium transport code {\it Smeagol}  and applied to the prototypical 
cases of benzene molecules sandwiched between gold electrodes. The self-interaction
corrected Kohn-Sham highest occupied molecular orbital now reproduces closely the negative of 
the molecular ionization potential and is moved away from the gold Fermi energy.
This leads to a drastic reduction of the low bias current in much better agreement with experiments.
\end{abstract}


\maketitle
Electronic transport in molecular devices is currently an area of much research interest.
Possible applications of such devices include high-performance computer components 
\cite{gates1}, chemical sensors \cite{chem}, disposable electronics, as well as possible 
medical applications \cite{virus}.

Recently it has become possible to construct single molecule junctions and measure 
their transport properties. Methods used to construct such devices include mechanically 
controllable break junctions \cite{Reed,Tsutsui,Ulrich}, STM tips \cite{Ulrich,Tao}, lithographically 
fabricated nanoelectrodes \cite{Ghosh}, and colloid solutions \cite{Dadosh}. However, there is much 
disagreement between the results for different experimental methods, with differences in the conductance 
of up to three orders of magnitude being reported for the same molecule \cite{Reed,Tsutsui,Tao,Dadosh}.  
For this reason {\it ab initio} quantum transport algorithms are becoming increasingly important.

Most computational methods for calculating electronic transport involve the combination of scattering 
theory in the form of the non-equilibrium Green's function (NEGF) formalism \cite{negf}, with an electronic 
structure method such as density-functional theory (DFT) \cite{dft,smeagol,negf-dft}. Other schemes include 
time-dependent DFT \cite{tddft} or many-body methods \cite{MB}. However, there is also much disagreement 
between the different theoretical methods, as well as between experiments and theory. In the prototypical 
case of benzenedithiol (BDT) attached to gold surfaces, the conductance for the most probable contact geometry
calculated with NEGF and DFT is higher than that of {\it any} experiments by at least one order 
of magnitude \cite{DiVentra,Ratner_1,Ratner_2,Ratner_3}. Still it is desirable to use DFT over
many-body approaches since it is conceptually simple and computationally both undemanding 
and scalable.
\begin{figure}[ht]
\begin{center}
\includegraphics[width=6.5cm,clip=true]{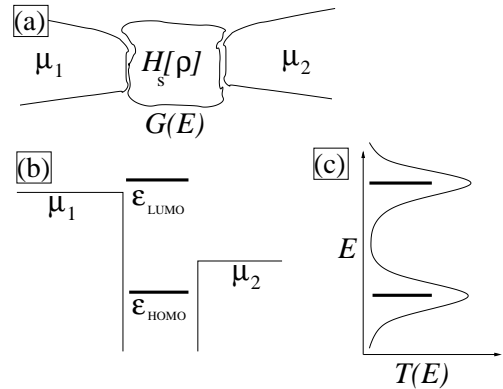}
\end{center}
\caption{\small{Schematic diagram of a metal-molecule-metal junction. (a) A scattering region 
is sandwiched between two current/voltage probes kept at the chemical potentials $\mu_1$ 
and $\mu_2$. (b) Energy level line up and (c) transmission coefficient as a function of energy.}}
\label{Fig1}
\end{figure}

A schematic illustration of the NEGF method is presented in figure \ref{Fig1}. A scattering region
is sandwiched between two current/voltage probes kept at different chemical potentials 
$\mu_{1/2}=E_\mathrm{F}\pm eV/2$ with $E_\mathrm{F}$ the Fermi energy, 
$V$ the bias and $e$ the electron charge. This is described by a Hamiltonian 
$H_\mathrm{s}$, which is used to construct the non-equilibrium Green's function $G(E)$
\begin{equation}
G(E)=\lim_{\eta\rightarrow 0}[(E+i\eta)-H_\mathrm{s}-\Sigma_\mathrm{1}-\Sigma_\mathrm{2}]^{-1}\:,
\label{negfmx}
\end{equation}
where $\Sigma_\mathrm{1/2}$ are the self-energy for the electrodes. $G(E)$ enters in a self-consistent 
procedure \cite{negf,smeagol,negf-dft} for evaluating the energy level line up (see Fig.\ref{Fig1}b) and hence
the two probe-current. This is simply the integral between $\mu_1$ and $\mu_2$ (the bias window) 
of the transmission coefficient $T(E)$,
which is a superposition of resonances located in correspondence to the molecular energy levels 
(the highest occupied molecular orbital - HOMO - and the lowest unoccupied molecular
orbital - LUMO - in Fig. \ref{Fig1}c).

The crucial point is that $H_\mathrm{s}$ and thus $G(E)$ are constructed from the 
single-particle states calculated by an associated electronic structure theory, which needs to meet 
several requirements \cite{tbsic}. First the single particle levels must closely resemble the physical 
removal energies of the system, i.e. the molecular levels and $E_\mathrm{F}$ should line up correctly. 
Secondly, it should work both at integer and fractional occupation, ensuring correct charging of the 
molecular levels. Finally the matrix elements between the leads and the molecule should be calculated accurately.

DFT in its Kohn-Sham (KS) formulation \cite{ks} is the most widely used electronic structure theory for transport,
although in principle it does not meet any of the previous requirements. In fact DFT observables are 
integral quantities of the KS eigenvalues (the total energy) or of the KS wave-functions (the 
charge density), with the individual KS orbitals having little meaning. There is however an important 
exception since the energy of the KS HOMO ($\epsilon_\mathrm{HOMO}$) is the negative of the ionization potential $I$ 
of the system \cite{janak,PPLB}. This suggests that, at least for moderate bias and HOMO transport, KS theory 
can be used for transport calculations \cite{note}. Unfortunately for standard local functionals,
such as the local density approximation (LDA), $\epsilon_\mathrm{HOMO}$ is nowhere near $-I$.

Most of the failures of LDA can be traced back to the self-interaction (SI), i.e. the 
spurious interaction of an electron with the Hartree and exchange-correlation (XC) potentials generated by itself \cite{PZ}. 
In the case of an exact exchange theory (for instance Hartree-Fock) the self-Hartree cancels with the self-XC
potential, however for LDA the cancellation is incomplete. This results in the eigenstates 
of a molecule being too high in energy, which translates in erroneously positioned peaks in the transmission 
coefficient. Self-interaction corrections (SIC) are possible \cite{PZ}, however their implementation in typical
solid state codes is cumbersome, since the theory become orbital dependent and the energy
minimization cannot follow the standard KS scheme. 

These problems become even more serious when SIC is combined with the NEGF method since 
the KS orbitals are never individually available. Recently we have implemented \cite{dasc} an alternative 
and approximated method for dealing with SIC. This is based on an atomic approximation (ASIC) \cite{pseudoSIC}, 
where the SIC orbitals that minimize the energy are assumed to be atomic-like. The correction thus 
becomes atomic and no information are needed other than the charge density and the ASIC projectors
\cite{dasc}. Importantly ASIC produces single-particle energy levels rather close to the experimental molecular removal 
energies, and in particular $\epsilon_\mathrm{HOMO}\approx -I$ and the $\epsilon_\mathrm{HOMO}$ for
negatively charged molecules is close to the molecular affinity. As an example ASIC places $\epsilon_\mathrm{HOMO}$
for 1,2-BDT at 8.47~eV to compare with the LDA value of 4.89~eV and the experimental ionization
potential $\sim$8.5~eV \cite{shen}.

We have numerically implemented the ASIC method \cite{dasc} in the localized atomic orbital code SIESTA \cite{siesta},
which is the DFT platform for our transport code {\it Smeagol} \cite{smeagol}, and carried out calculations for the 
prototypical Au/Benzene/Au molecular devices. We use a double zeta polarized basis set 
\cite{siesta} for carbon and sulfur $s$ and $p$ orbitals, double zeta for the 
1$s$ orbitals of hydrogen and 6$s$-only double zeta for gold. The mesh cut-off is 200 Ry 
and we consider 500 real and 80 complex energy points for integrating the Green's function. 

\begin{figure}[ht]
\begin{center}
\includegraphics[width=7.5cm,clip=true]{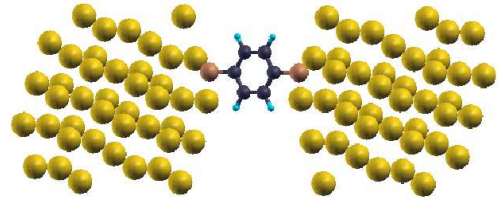}
\end{center}
\caption{\small{BDT molecule attached to the hollow site of the Au (111) surface. 
The sulfur-surface distance is 1.9\AA. Color code: Au=yellow, C=black, S=dark yellow, H=blue.}}
\label{Fig2}
\end{figure}

First we consider a 1,4-BDT molecule sandwiched between two fcc (111) gold surfaces and
attached to the gold hollow site (figure \ref{Fig2}). We optimize
the distance between the sulfur atom and the gold surface to a value of 1.9\AA, which agrees with 
previous calculations \cite{Ratner_1,AuSdist}. 

\begin{figure}[ht]
\begin{center}
\includegraphics[width=9.0cm,clip=true]{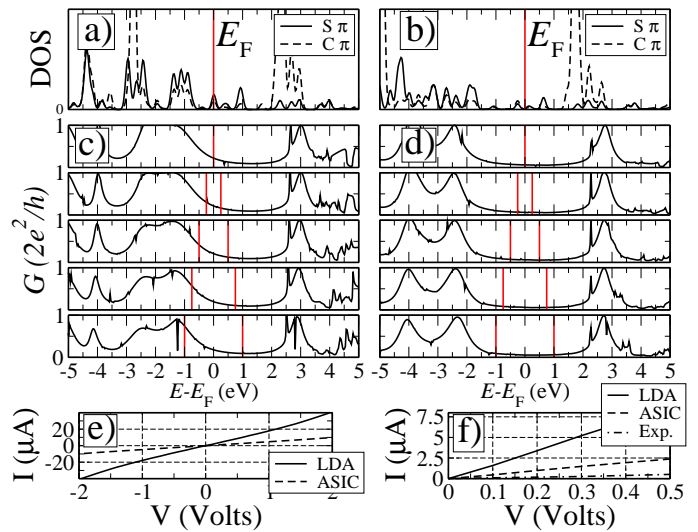}
\end{center}
\caption{\small{Transport properties of a BDT molecule attached to the gold (111) hollow site. The left
plots correspond to LDA and the right ones to ASIC. The upper panels are the DOS of the S 
and C $\pi$ orbitals ((a) and (b)), the middle are the transmission coefficients as a function 
of energy for various bias ((c) and (d)) and the lower are the $I$-$V$ curves. Figure (f) is a zoom of
(e) and compares our results with experiments from reference \cite{Tao}. The vertical lines in
(c) and (d) mark the bias window.}}
\label{Fig3}
\end{figure}

The orbital resolved density of states (DOS), transmission coefficient and $I$-$V$ curves are presented
in figure \ref{Fig3} for both LDA and ASIC. From the DOS (panels (a) and (b)) it is clear 
that the effect of ASIC is that of shifting the occupied orbitals downwards relative to $E_\mathrm{F}$ of the gold. 
The HOMO-LUMO gap is considerably larger than that of the LDA, and most importantly in the case of ASIC there 
is little DOS originating from the molecule at $E_\mathrm{F}$.
This has profound effects over the electron transmission. The LDA peaks of $T(E)$
arising from occupied orbitals are shifted downwards in energy and away from $E_\mathrm{F}$. 
At variance from LDA (Fig. \ref{Fig3}c), where $T(E_\mathrm{F})$ is dominated 
by a resonance at $\epsilon_\mathrm{HOMO}$, the ASIC transmission (Fig. \ref{Fig3}d) is through the BDT gap
and therefore it is tunneling-like. This results in a drastic reduction of the low-bias current when going from
LDA to ASIC (Fig. \ref{Fig3}e). The ASIC-calculated conductance at zero bias is now about 0.06$G_0$ ($G_0=2e^2/h$), 
compared to 0.23$G_0$ of LDA. A conductance of 0.06$G_0$ is much closer to the value of 
0.011$G_0$ obtained by Xiao et. al. \cite{Tao} and is actually lower than values 0.09-0.14$G_0$ 
obtained by Tsutsui et. al. \cite{Tsutsui}.

Other anchoring configurations to the (111) surface were investigated
and their zero-bias conductance are shown in Table \ref{Tab}. Note that ASIC returns 
values in the region of about 0.06$G_0$ for several different anchoring geometries. 
Such stability with variation of the anchoring structure is important as the peaks in the experimental 
conductance histograms are relatively sharp \cite{Tao}. This indicates that the different metal-molecule 
junctions have similar conductances, despite the fact that the anchoring of the molecule to the surface/tip may vary.
\begin{table}[ht]
\begin{tabular}{lccc}
\hline\hline
Anchoring & $d$ (\AA) & $G_\mathrm{LDA}$ ($G_0$) & $G_\mathrm{ASIC}$ ($G_0$) \\ \hline\hline
Ho & 1.9 & 0.23 & 0.06\\ 
Ho & 1.8 & 0.16 & 0.05 \\ 
Ho & 2.1 & 0.32 & 0.07 \\ 
Ho & 2.5 & 0.77 & 0.14 \\ 
Ho (30$^{\circ}$) & 1.9 & 0.18 & 0.04 \\ 
Br & 2.09 & 0.11 & 0.06 \\ 
Ad & 2.39 & 0.11 & 0.10 \\ 
Asy Ho & 1.9/2.3 & 0.33 & 0.06 \\ 
Ho/Ad & 1.9/2.39 & 0.35 & 0.03 \\ \hline\hline
\end{tabular}
\caption{\label{Tab}Zero-bias conductance for different configurations of BDT on gold (111). 
Experimental values include 0.011$G_0$ \cite{Tao} and 0.09$G_0$ \cite{Ulrich}. 
The anchoring configurations investigated are: hollow site (Ho), bridge site (Br), Au adatom (Ad). 
Ho (30$^{\circ}$) describes a hollow site with BDT at a 30$^{\circ}$ angle
with respect to the transport direction, and the two last rows correspond to asymmetric 
anchoring to the two electrodes. $d$ is distance between the S atom of the thiol group and the plane 
of the gold surface (or to the adatom for Ad).}
\end{table}

\begin{figure}[ht]
\begin{center}
\includegraphics[width=7.5cm,clip=true]{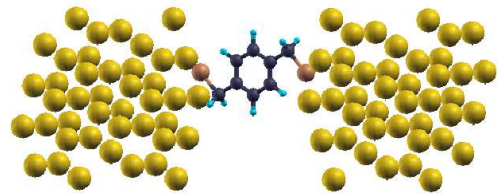}
\end{center}
\caption{\small{BDMT molecule attached to the hollow sites of the gold (111) surface. The contact 
geometry is identical to that for BDT.  Color code: Au=yellow, C=black, S=dark yellow, H=blue.}}
\label{Fig4}
\end{figure}

As a second case we investigate benzenedimethanethiol (BDMT) molecules on the same
gold surface (Fig. \ref{Fig4}). Also for this molecule we consider a number of anchoring 
structures, although here we report only for the hollow site since different structures do not
present qualitatively different results. 

\begin{figure}[ht]
\begin{center}
\includegraphics[width=9.0cm,clip=true]{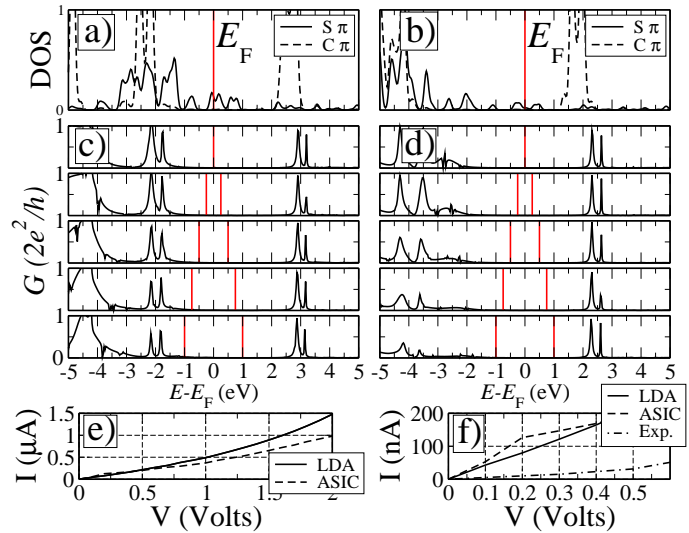}
\end{center}
\caption{\small{Transport properties of a BDMT molecule attached to the gold (111) hollow site. The left
panels correspond to LDA and the right ones to ASIC. The upper panels are the DOS of the S 
and C $\pi$ orbitals ((a) and (b)), the middle are the transmission coefficients as a function 
of energy for various bias ((c) and (d)) and the lower are the $I$-$V$ curves. Figure (f) is a zoom of
(e) and compares our results with experiments from reference \cite{Tao}. The vertical lines in
(c) and (d) mark the bias window.}}
\label{Fig5}
\end{figure}

In contrast to BDT, the DOS of BDMT already presents a large HOMO-LUMO gap in LDA
(see Fig. \ref{Fig5}a), which is further increased by the ASIC (\ref{Fig5}b).
This time ASIC has the only effect of changing the alignment of $\epsilon_\mathrm{HOMO}$ with 
respect to $E_\mathrm{F}$, which cuts through the HOMO-LUMO gap for
both LDA and ASIC. Therefore both LDA and ASIC offer a picture of tunneling-like transport 
through the molecular gap, and little difference can be found (see Fig. \ref{Fig5}c, d, e and f). 
Hence, the $I$-$V$ curves are quite similar, with currents approximately one
order of magnitude smaller than those of BDT. 

With these results in hand we can conclude that in general ASIC drastically improves the
agreement between theory and experiments. However some disagreement still remains. In particular it 
appears that even in the case of ASIC the current at low bias is larger than that typically measured. 
Although an exhaustive comparison is
complicated by the fact that the experimental spread of the data is large, here we speculate
on the possible source of such disagreement. First one may argue that the contact geometry is not correct. Indeed recent
X-ray standing wave experiments \cite{adatom_thiol} demonstrate that S atoms in thiol groups on gold join more 
favorably to adatoms. This means that the Au-S-molecule moiety may be the one relevant
for the transport experiments. However calculations with two Au adatoms as anchoring sites lead to strong pinning
of $\epsilon_\mathrm{HOMO}$ to the gold $E_\mathrm{F}$ \cite{Ratner_1,Ratner_3}. ASIC does not change this
feature and the conductance remains large. An asymmetric anchoring configuration with one hollow site and 
one adatom (see last column of table \ref{Tab}) gives us a ASIC conductance of 0.03~$G_0$, still higher than
experiments. Note that the presence of the mobile Au-S-molecule moieties seems also difficult to 
conciliate with the relative robustness of the peaks in the conductance histograms of breaking-junctions
\cite{Tao}.

Hence we perhaps have to accept the fact that the disagreement between theory and experiments persists.
Notably the problem is now that of calculating accurately tunneling matrix elements between the S and the surface. 
The problem is thus critically dependent on the quality of the wave-function and in turn of the
actual scattering potential, for which ASIC does not offer substantial improvement over the LDA. In particular
ASIC still overestimates the polarizability of molecules \cite{ASICPOL}, with a quantitatively incorrect
prediction of the response exchange and correlation field. In addition it is important to remark that 
we have applied the ASIC only to the molecular degrees of freedom, without correcting the Au atoms. 
It is thus likely that the Au 6$s$ orbitals at the surface are too extended, leading to a larger current.

In conclusion, we have demonstrated that a simple SIC scheme is able of lowering the energy levels of the 
occupied molecular orbitals, which now resemble closely the actual vertical removal energies. This has
profound consequences over the transport properties of metal/molecule/metal junctions since 
spurious resonances at the Fermi level can be removed, leading to tunneling transport for molecules
for which LDA erroneously predicts metallic conductance. The agreement with experiments is thus 
greatly improved.

We thank Alex~Reily Rocha, Chaitanya~Pemmaraju, Kieron~Burke and Alessio~Filippetti for useful discussions. This work is funded 
by Science Foundation of Ireland (grant SFI02/IN1/I175). Computational resources have been provided by the 
HEA IITAC project managed by the Trinity Center for High Performance Computing and by
ICHEC.
 \\

\end{document}